\documentclass[preprint,prd,aps,showpacs,showkeys,nofootinbib]{revtex4}
\usepackage{graphicx}
\begin{document}
\title{The two-loop corrections to lepton MDMs and EDMs in the EBLMSSM}
\author{Xing-Xing Dong$^{1}$\footnote{dxx$\_$0304@163.com},
Shu-Min Zhao$^{1}$\footnote{zhaosm@hbu.edu.cn},
Hai-Bin Zhang$^{1}$\footnote{hbzhang@hbu.edu.cn},
Tai-Fu Feng$^{1}$\footnote{fengtf@hbu.edu.cn}}
\affiliation{$^1$ Department of Physics, Hebei University, Baoding 071002,
China}
\begin{abstract}
Extending BLMSSM with exotic Higgs superfields $(\Phi_{NL},\varphi_{NL})$ and superfields ($Y,Y^\prime$), one obtains the new model called as EBLMSSM, where exotic leptons are heavy and have tree level couplings with SM lepton. In this model, some new parameters with CP-violating phases are considered, so there are new contributions to lepton anomalous magnetic dipole moments (MDMs) and electric dipole moments (EDMs). Therefore, we study the one-loop, two-loop Barr-Zee and two-loop Rainbow type corrections to lepton MDMs and EDMs in the EBLMSSM. Considering the constraints from the lightest CP-even Higgs mass and decays, we calculate the corresponding numerical results. In our used parameter space, the new physics contributions to lepton MDMs are large, which can remedy the deviation between the SM prediction and experimental result well. New introduced CP-violating phases also affect the lepton EDMs in a certain degree.
\end{abstract}


\maketitle

\section{Introduction}
Combined with the experimental datas of the ATLAS\cite{h0ATLAS} and CMS\cite{h0CMS} Collaborations, the scientists released a Higgs boson detected on the Large Hadron Collider (LHC), and its mass is $m_{h^0}=125.18\pm0.16\rm {GeV}$\cite{PDG2018}. As a basic particle predicted by the standard model (SM), the discovery of Higgs boson has made the SM a great success. Since Schwinger first proposed the electron MDM, it has been recognized that the magnetic dipole moment of lepton can provide accurate testing of Quantum Electrodynamics (QED) and subsequently of the SM\cite{S.eMDM}. It will be a very meaningful observable to study the lepton MDMs.

Although the contribution from QED $a_l^{QED}$ plays a major role in the lepton MDMs, it is not the only factor. The contribution of hadron $a_l^{HAD}$ is very important, whose corrections come from hadronic vacuum polarization (VP) and hadronic light-by-light (HLbL) scattering contributions. In addition, although the weak interaction $a_l^{EW}$ is inhibited by the weak gauge boson mass, it also has a certain influence on the lepton MDMs. Therefore, the contributions from lepton MDMs in SM can be expressed as\cite{ae,expau1,SMau}:
\begin{eqnarray}
a_l^{SM}=a_l^{QED}+a_l^{EW}+a_l^{HAD}.
\end{eqnarray}

In 1947, electron MDM was discovered in an atomic physics experiment\cite{expae}. Since then, scientists have continuously improved the measurement accuracy of lepton MDMs, and the corresponding theoretical calculation with great precision has also been carried out within the SM. The authors of the reference\cite{Csexpae} give the most accurate value of the fine structure up to now by determining the mass of the Cs atom. Combining the corresponding theoretical values of the SM\cite{ae}, one can derive a $2.4\sigma$ discrepancy in the electron MDM\cite{Csexpae,deltaae}:
\begin{eqnarray}
\Delta a_{e} =a^{exp}_e-a^{SM}_{e}=(-88\pm36)\times 10^{-14}.\label{deltaae}
\end{eqnarray}

On the other hand, the muon MDM has $3.7\sigma$ deviation between experiment and theory\cite{expau1,expau2,deltaau1,deltaau2}:
\begin{eqnarray}
\Delta a_{\mu} =a^{exp}_{\mu}-a^{SM}_{\mu}=(274\pm73)\times 10^{-11}.\label{deltaau}
\end{eqnarray}
Most crucially, the sign of $\Delta a_{\mu}$ is opposite to that of $\Delta a_{e}$. It is worth noting that if there is no flavor in the lepton zone, the muon MDM and electron MDM are subject to the lepton mass scaling, thus expecting the same sign deviation. The deviation of muon MDM is the same as the order of weak correction, so it can naturally be explained by weak-scale physics, but electron MDM cannot be reasonably explained for its negative sign. The existence of muon MDM and electron MDM and their opposite sign indicate that there may be new physics contributions beyond SM.

In 1964, Cronin and Fitch discovered the charge conjugate and parity (CP)-violating decays of the $K$ meson\cite{CPviolation}. As the physical quantities for probing sources of CP violation, the EDMs of lepton are researched. The present experiments have reported that the upper bound of electron EDM is $|d_e|<8.7\times10^{-29}$ e.cm\cite{deACME,PDG2018,ACME}, the muon and tau EDMs are respectively $|d_{\mu}|<1.9\times10^{-19}$ e.cm and $|d_{\tau}|<1\times10^{-17}$ e.cm\cite{PDG2018,dudt}. In order to explain the observed CP-violating effects, the CP-violating source is artificially placed in the SM. However, the theoretical predictions for lepton EDMs in the SM are tiny, such as the electron EDM is around $10^{-38}$ e.cm\cite{SMde1,SMde2,SMde3}, which is too small to be detected by the present experiment. Therefore, the origin and mechanism of CP-violation are still not well explained. Scientists are trying to find CP-violating phases in the new physics beyond SM to better explain the CP-violating mechanism\cite{NPdl1,NPdl2,NPdl3,NPdl4,NPdl5,NPdl6,BLMDM}.

Physicists have established many new models. The minimal supersymmetric extension of the standard model (MSSM)\cite{MSSM1,MSSM2,MSSM3,MSSM4} is one of the most attractive candidates. To explain asymmetry of matter-antimatter in the universe and the neutrino quality issues, baryon number (B) and lepton number (L) need to be considered even broken at TeV scale. Then the BLMSSM\cite{BLMSSM1,BLMSSM2,BLMSSM3,BLMSSM4} is obtained, which is a simple extension of the MSSM with local gauged B and L. Although BLMSSM can explain many problems well, the quality of exotic leptons is not heavy enough due to the small values of parameters ($Y_{e4},~Y_{e5},~v_u$ and $~v_d$). The masses of the exotic leptons are around 100GeV, which may be ruled out easily by the future experiments. This is a very fatal flaw for BLMSSM and even relates to whether it will exist. Therefore, we hope to introduce two exotic Higgs superfields $\Phi_{NL}$ and $\varphi_{NL}$ to the BLMSSM, so that exotic leptons can become heavy enough. Besides, the superfields $Y$ and $Y^\prime$ are taken into account to make heavy exotic leptons unstable. Not only that, a mix between the fourth and
fifth generation leptons is also considered. Therefore, we obtain an extended model of BLMSSM, which is called EBLMSSM\cite{EBL1,EBL2}. In this model, we introduce new superfields, new interactional terms, new mass matrices of particles and new CP-violating phases. In the following, we calculate the lepton MDMs and EDMs at one-loop and two-loop level using the effective Lagrangian method, and receive the concrete numerical results that will coincide the present experiment dates well.

After this introduction, we introduce the contents of EBLMSSM briefly. The needed mass matrices and couplings are given out in Section II. In Section III, we deduce the one-loop, two-loop Barr-Zee and two-loop Rainbow type corrections to lepton MDMs and EDMs in detail. The corresponding numerical results are discussed in Section IV. The last Section is devoted to our conclusion. Some of the two-loop results will be placed in the Appendix A.
\section{the EBLMSSM}
We extend BLMSSM with the superfields $\Phi_{NL},\varphi_{NL},Y,Y'$ and obtain EBLMSSM \cite{EBL1}. Same as BLMSSM, the local gauge
group of the EBLMSSM is $SU(3)_{C}\otimes SU(2)_{L}\otimes U(1)_{Y}\otimes U(1)_{B}\otimes U(1)_{L}$ \cite{BLMSSM1,EBL1,group1,group2}. In BLMSSM, the exotic leptons are not heavy enough and may be excluded by the future
experiments. The superfields $\Phi_{NL},\varphi_{NL}$ added in the EBLMSSM affect the exotic lepton masses and make them heavy. On the other hand, heavy particles should decay quickly, so the superfields $Y,Y'$ are introduced.
The lightest mass eigenstate of $Y$ and $Y'$ mixing
can be considered as a new dark matter candidate.

In EBLMSSM, the superfields beyond BLMSSM are given out in the TABLE \ref{superfields}. The superfields in BLMSSM\cite{BLsuperfield1,BLsuperfield2} are not shown here for saving space.
\begin{table}[t]
\caption{ The superfields beyond BLMSSM }
\begin{tabular}{|c|c|c|c|c|c|}
\hline
Superfields & $SU(3)_C$ & $SU(2)_L$ & $U(1)_Y$ & $U(1)_B$ & $U(1)_L$\\
\hline
$\hat{\Phi}_{NL}$ & 1 & 1 & 0 & 0 & -3 \\
\hline
$\hat{\varphi}_{NL}$ & 1 & 1 & 0 & 0 & 3\\
\hline
$Y$ & 1 & 1 & 0 & 0 & $2+L_4$
\\ \hline
$Y'$ & 1 & 1 & 0 & 0 & $-(2+L_4)$ \\
\hline
\end{tabular}
\label{superfields}
\end{table}

The superpotential of EBLMSSM reads as
\begin{eqnarray}
&&{\cal W}_{{EBLMSSM}}={\cal W}_{{MSSM}}+{\cal W}_{B}+{\cal W}_{L}+{\cal W}_{X}+{\cal W}_{Y}\;,
\nonumber\\
&&{\cal W}_{L}=\lambda_{L}\hat{L}_{4}\hat{L}_{5}^c\hat{\varphi}_{NL}+\lambda_{E}\hat{E}_{4}^c\hat{E}_{5}
\hat{\Phi}_{NL}+\lambda_{NL}\hat{N}_{4}^c\hat{N}_{5}\hat{\Phi}_{NL}
+\mu_{NL}\hat{\Phi}_{NL}\hat{\varphi}_{NL}\nonumber\\&&\hspace{1.2cm}+Y_{{e_4}}\hat{L}_{4}\hat{H}_{d}\hat{E}_{4}^c+Y_{{\nu_4}}\hat{L}_{4}\hat{H}_{u}\hat{N}_{4}^c
+Y_{{e_5}}\hat{L}_{5}^c\hat{H}_{u}\hat{E}_{5}+Y_{{\nu_5}}\hat{L}_{5}^c\hat{H}_{d}\hat{N}_{5}
\nonumber\\
&&\hspace{1.2cm}
+Y_{\nu}\hat{L}\hat{H}_{u}\hat{N}^c+\lambda_{{N^c}}\hat{N}^c\hat{N}^c\hat{\varphi}_{L}
+\mu_{L}\hat{\Phi}_{L}\hat{\varphi}_{L}\;,
\nonumber\\&&
{\cal W}_{Y}=\lambda_4\hat{L}\hat{L}_{5}^c\hat{Y}+\lambda_5\hat{N}^c\hat{N}_{5}\hat{Y}^\prime
+\lambda_6\hat{E}^c\hat{E}_{5}\hat{Y}^\prime+\mu_{Y}\hat{Y}\hat{Y}^\prime,\label{superpotential}
\end{eqnarray}
where ${\cal W}_{{MSSM}}$ represents the superpotential of MSSM.
${\cal W}_{B}$ and ${\cal W}_{X}$ denote the corresponding terms originating from BLMSSM\cite{BLsuperfield1}. Compared with BLMSSM, ${\cal W}_{Y}$ is the new part, which include the new effects from $Y$-lepton-exotic lepton and $\tilde{Y}$-slepton-exotic slepton couplings to lepton MDMs and EDMs. Additionally, $\lambda_4(\lambda_6)$ is the coupling coefficient of $Y$-lepton-exotic lepton and $\tilde{Y}$-lepton-exotic slepton couplings. In our previous work\cite{EBL1,EBL2}, $\lambda_4^2(\lambda_6^2)$ is considered as a $3\times 3$ matrix. Only the diagonal elements($(\lambda_4^2)_{II}=(\lambda_6^2)_{II}=(Lm^2)_{II}$, $I$ represents the $I$-th generation charged lepton) have contributions to the MDMs and EDMs of lepton. ${\cal W}_{L}$ also possesses new contents beyond BLMSSM. These new contents influence the masses of exotic lepton, exotic neutrino, exotic slepton and lepton neutralino.

The EBLMSSM soft breaking terms can be found in our previous work\cite{EBL1,EBL2}. The $SU(2)_L$ doublets are $H_{u}$ and $H_{d}$, whose nonzero vacuum expectation values (VEVs) are $\upsilon_{u}$ and $\upsilon_{d}$.
\begin{eqnarray}
&&H_{u}=\left(\begin{array}{c}H_{u}^+\\{1\over\sqrt{2}}\Big(\upsilon_{u}+H_{u}^0+iP_{u}^0\Big)\end{array}\right),
~~~~~~
H_{d}=\left(\begin{array}{c}{1\over\sqrt{2}}\Big(\upsilon_{d}+H_{d}^0+iP_{d}^0\Big)\\H_{d}^-\end{array}\right).
\end{eqnarray}
The $SU(2)_L$ singlets $\Phi_{L},\varphi_{L},\Phi_{NL},\varphi_{NL}$
obtain the nonzero VEVs $\upsilon_{{L}},\overline{\upsilon}_{{L}},\upsilon_{{NL}},\overline{\upsilon}_{{NL}}$ respectively, which are shown here
\begin{eqnarray}
&&\Phi_{L}={1\over\sqrt{2}}\Big(\upsilon_{L}+\Phi_{L}^0+iP_{L}^0\Big),~~~~~~~~~~~~
\varphi_{L}={1\over\sqrt{2}}\Big(\overline{\upsilon}_{L}+\varphi_{L}^0+i\overline{P}_{L}^0\Big),
\nonumber\\
&&\Phi_{NL}={1\over\sqrt{2}}\Big(\upsilon_{NL}+\Phi_{NL}^0+iP_{NL}^0\Big),~~~~
\varphi_{NL}={1\over\sqrt{2}}\Big(\overline{\upsilon}_{NL}+\varphi_{NL}^0+i\overline{P}_{NL}^0\Big).
\end{eqnarray}
We define that the parameters $\tan\beta=\upsilon_u/\upsilon_d,\tan\beta_L=\bar{\upsilon}_L/\upsilon_L$ and
$\tan\beta_{NL}=\bar{\upsilon}_{NL}/\upsilon_{NL}$.

In EBLMSSM, the contributions to the lepton MDMs and EDMs are affected by corrected particles, such as slepton, sneutrino, exotic lepton, exotic neutrino, exotic slepton, lepton neutralino, $Y$ and $\tilde{Y}$. We will discuss these particles in detail. The Lagrangian of exotic lepton mass matrix in EBLMSSM is shown here.
\begin{eqnarray}
&&-{\cal L}_{{L^\prime}}^{mass}=\left(\begin{array}{ll}\bar{e}_{{4R}},&\bar{e}_{{5R}}\end{array}\right)
\left(\begin{array}{ll}-{1\over\sqrt{2}}\lambda_{L}\overline{\upsilon}_{NL},&{1\over\sqrt{2}}Y_{{e_5}}\upsilon_{u}\\
-{1\over\sqrt{2}}Y_{{e_4}}\upsilon_{d},&{1\over\sqrt{2}}\lambda_{E}\upsilon_{NL}
\end{array}\right)\left(\begin{array}{l}e_{{4L}}\\e_{{5L}}\end{array}\right)+h.c.
\label{exotic lepton}
\end{eqnarray}
The exotic lepton masses are heavier than those in BLMSSM, the reason is that the diagonal elements in Eq.(\ref{exotic lepton}) include $\upsilon_{NL}$ and $\bar{\upsilon}_{NL}$, which can be large parameters.
To obtain mass eigenstates, we use the unitary transformations
\begin{eqnarray}
&&\left(\begin{array}{l}e_{{4L}}^\prime\\e_{{5L}}^\prime\end{array}\right)
=U_{{L}}^\dagger\cdot\left(\begin{array}{l}e_{{4L}}\\e_{{5L}}\end{array}\right)\;,\;\;
\left(\begin{array}{l}e_{{4R}}^\prime\\e_{{5R}}^\prime\end{array}\right)
=W_{{L}}^\dagger\cdot\left(\begin{array}{l}e_{{4R}}\\e_{{5R}}\end{array}\right).
\label{Qmixing-2/3-a}
\end{eqnarray}

Similar as the exotic lepton condition, heavy exotic neutrinos are also gotten through the following Lagrangian:
\begin{eqnarray}
&&-{\cal L}_{{N^\prime}}^{mass}=\left(\begin{array}{ll}\bar{\nu}_{{4R}}^\prime,&\bar{\nu}_{{5R}}^\prime\end{array}\right)
\left(\begin{array}{ll}{1\over\sqrt{2}}\lambda_{L}\overline{\upsilon}_{NL},&-{1\over\sqrt{2}}Y_{{\nu_5}}\upsilon_{d}\\
{1\over\sqrt{2}}Y_{{\nu_4}}\upsilon_{u},&{1\over\sqrt{2}}\lambda_{NL}\upsilon_{NL}
\end{array}\right)\left(\begin{array}{l}\nu_{{4L}}^\prime\\\nu_{{5L}}^\prime\end{array}\right)+h.c.
\label{Nmass-matrix}
\end{eqnarray}

Being different from BLMSSM, the exotic sleptons of 4 generation and 5 generation in EBLMSSM have mix and their mass squared matrix is $4\times4$.
The elements of exotic slepton
mass matrix $\mathcal{M}^2_{\tilde{E}}$ are deduced as follows
\begin{eqnarray}
&&\mathcal{M}^2_{\tilde{E}}(\tilde{e}_5^{c*}\tilde{e}_5^{c})=
\lambda_L^2\frac{\bar{\upsilon}_{NL}^2}{2}+\frac{\upsilon_u^2}{2}|Y_{e_5}|^2+M^2_{\tilde{L}_5}
-\frac{g_1^2-g_2^2}{8}(\upsilon_d^2-\upsilon_u^2)-g_L^2(3+L_4)V_L^2,
\nonumber\\&&\mathcal{M}^2_{\tilde{E}}(\tilde{e}_5^{*}\tilde{e}_5)=\lambda_E^2\frac{\upsilon_{NL}^2}{2}
 +\frac{\upsilon_u^2}{2}|Y_{e_5}|^2+M^2_{\tilde{e}_5}+\frac{g_1^2}{4}(\upsilon_d^2-\upsilon_u^2)
 +g_L^2(3+L_4)V_L^2,
\nonumber\\&&\mathcal{M}^2_{\tilde{E}}(\tilde{e}_4^{*}\tilde{e}_4)=\lambda_L^2\frac{\bar{\upsilon}_{NL}^2}{2}
+\frac{g_1^2-g_2^2}{8}(\upsilon_d^2-\upsilon_u^2)+\frac{\upsilon_d^2}{2}|Y_{e_4}|^2+M^2_{\tilde{L}_4}
+g_L^2L_4V_L^2,
\nonumber\\&&\mathcal{M}^2_{\tilde{E}}(\tilde{e}_4^{c*}\tilde{e}_4^{c})=
\lambda_E^2\frac{\upsilon_{NL}^2}{2}-\frac{g_1^2}{4}(\upsilon_d^2-\upsilon_u^2)+\frac{\upsilon_d^2}{2}|Y_{e_4}|^2+M^2_{\tilde{e}_4}
-g_L^2L_4V_L^2,
\nonumber\\&&\mathcal{M}^2_{\tilde{E}}(\tilde{e}_4^{*}\tilde{e}_5)
=\upsilon_dY_{e_4}^*\lambda_E\frac{\upsilon_{NL}}{2}+\lambda_LY_{e_5}\frac{\bar{\upsilon}_{NL}v_u}{2},
~~~\mathcal{M}^2_{\tilde{E}}(\tilde{e}_5\tilde{e}_5^{c})=\mu^*\frac{\upsilon_d}{\sqrt{2}}Y_{e_5}+A_{e_5}\frac{\upsilon_u}{\sqrt{2}},
\nonumber\\&&\mathcal{M}^2_{\tilde{E}}(\tilde{e}_4^{c}\tilde{e}_5)=\mu_{NL}^*\lambda_E
\frac{\bar{\upsilon}_{NL}}{\sqrt{2}}-A_{LE}\lambda_E\frac{\upsilon_{NL}}{\sqrt{2}},
~~~\mathcal{M}^2_{\tilde{E}}(\tilde{e}_4\tilde{e}_5^{c})=-\mu_{NL}^*\frac{\upsilon_{NL}}{\sqrt{2}}\lambda_L-A_{LL}\lambda_L\frac{\bar{\upsilon}_{NL}}{\sqrt{2}},
\nonumber\\&&\mathcal{M}^2_{\tilde{E}}(\tilde{e}_4\tilde{e}_4^{c})=\mu^*\frac{\upsilon_u}{\sqrt{2}}Y_{e_4}+A_{e_4}\frac{\upsilon_d}{\sqrt{2}},
~~~~~~~~\mathcal{M}^2_{\tilde{E}}(\tilde{e}_5^{c}\tilde{e}_4^{c*})
=Y_{e_5}\lambda_E\frac{\upsilon_u\upsilon_{NL}}{2}-\lambda_LY_{e_4}^*\frac{\bar{\upsilon}_{NL}v_d}{2},
\end{eqnarray}
where $V_L^2=\overline{\upsilon}^2_L-\upsilon^2_L+\frac{3}{2}(\overline{\upsilon}^2_{NL}-\upsilon^2_{NL})$.
In the base $(\tilde{e}_4,\tilde{e}_4^{c*},\tilde{e}_5,\tilde{e}_5^{c*})$, we diagonalize $\mathcal{M}^2_{\tilde{E}}$ by the matrix $Z_{\tilde{E}}$ through the formula
$Z^{\dag}_{\tilde{E}}\mathcal{M}^2_{\tilde{E}} Z_{\tilde{E}}=diag(m^2_{\tilde{E}^1},m^2_{\tilde{E}^2},m^2_{\tilde{E}^3},m^2_{\tilde{E}^4})$.

The mass squared matrix for the mix of $Y$ and $Y'$ is shown in the following form and diagonalized by the rotation matrix $Z_{Y}$
  \begin{eqnarray}
Z^{\dag}_{Y}\left(     \begin{array}{cc}
  |\mu_{Y}|^2+S_{Y} &-\mu_{Y}B_{Y} \\
    -\mu^*_{Y}B^*_{Y} & |\mu_{Y}|^2-S_{Y}\\
    \end{array}\right)  Z_{Y}=\left(     \begin{array}{cc}
 m_{{Y_1}}^2 &0 \\
    0 & m_{{Y_2}}^2\\
    \end{array}\right),
   ~~~~~\left(     \begin{array}{c}
  Y_{1} \\  Y_{2}\\
    \end{array}\right) =Z_{Y}^{\dag}\left( \begin{array}{c}
  Y \\  Y'^*\\
    \end{array}\right).
   \end{eqnarray}
Here $S_{Y}=g_{L}^2(2+L_{4})V_L^2$.

In EBLMSSM, there is a four-component Dirac spinor $\tilde{Y}$ made up of the superpartners of $Y$ and $Y'$,
\begin{eqnarray}
  &&-\mathcal{L}^{mass}_{\tilde{Y}}=\mu_Y\bar{\tilde{Y}}\tilde{Y}
  ,~~~~~~~~~~~~~~~~\tilde{Y} =\left( \begin{array}{c}
  \psi_{Y'} \\  \bar{\psi}_{Y}\\
    \end{array}\right).
   \end{eqnarray}

In the base $(i\lambda_L,\psi_{\Phi_L},\psi_{\varphi_L},\psi_{\Phi_{NL}},\psi_{\varphi_{NL}})$, we deduce the mass matrix of lepton neutralino, which can be diagonalized by the rotation matrix $Z_{NL}$.
\begin{equation}
\mathcal{M}_L=\left(     \begin{array}{ccccc}
  2M_L &2\upsilon_Lg_L &-2\bar{\upsilon}_Lg_L&3\upsilon_{NL}g_L &-3\bar{\upsilon}_{NL}g_L\\
   2\upsilon_Lg_L & 0 &-\mu_L& 0 & 0\\
   -2\bar{\upsilon}_Lg_L&-\mu_L &0& 0 & 0\\
   3\upsilon_{NL}g_L & 0 & 0 & 0 & -\mu_{NL}\\
   -3\bar{\upsilon}_{NL}g_L& 0&0&-\mu_{NL}&0
    \end{array}\right).
\end{equation}

In the EBLMSSM, with superpotential $\mathcal{W}_Y$ in Eq.(\ref{superpotential}), we deduce the tree-level coupling for lepton-exotic lepton-$Y$
\begin{eqnarray}
&&\mathcal{L}_{lL'Y}=\bar{l^I}\Big(\lambda_4W_L^{1i}Z_Y^{1j*}\omega_+ -\lambda_6U_L^{2i}Z_Y^{2j*}\omega_-\Big)L'_{i+3}Y_j^*+h.c.
\end{eqnarray}
The coupling for lepton-exotic slepton-$\tilde{Y}$ is also obtained
\begin{eqnarray}
&&\mathcal{L}_{l\tilde{E}\tilde{Y}}=\bar{\tilde{Y}}\Big(\lambda_4Z_{\tilde{E}}^{4i*}\omega_- -\lambda_6Z_{\tilde{E}}^{3i*}\omega_+\Big)l^I\tilde{E}_i^*+h.c.
\end{eqnarray}
Through $ig\sqrt{2}T^a_{ij}(\lambda^a\psi_jA_i^*-\bar{\lambda}^a\bar{\psi}_iA_j)$, we deduce the lepton-lepton neutralino-slepton coupling
\begin{eqnarray}
&&\mathcal{L}_{l\chi_{NL}^0\tilde{L}}=\sqrt{2}g_L\bar{\chi}_{L_j}^0\Big(Z_{N_L}^{1j}Z_{L}^{Ii}\omega_--Z_{N_L}^{1j*}Z_{L}^{(I+3)i*}\omega_+\Big)l^I\tilde{L}_i^+.
\end{eqnarray}
Similar as the $Z$ gauge boson, the $U(1)_L$ gauge boson $Z_L^{\mu}$ also has coupling with lepton
\begin{eqnarray}
\mathcal{L}_{Z_L^{\mu}ll}=-g_LZ^\mu_L\bar{l}^I\gamma_\mu l^I.
\end{eqnarray}

\section{The corrections to lepton MDMs and EDMs in the EBLMSSM}
The effective Lagrangian used here for the lepton MDMs and MDMs are given out as follows
\begin{eqnarray}
&&{\cal L}_{MDM}={e\over4m_{l}}\;a_{l}\;\bar{l}\sigma^{\mu\nu}
l\;F_{{\mu\nu}},\label{adm}
~~~~~~{\cal L}_{EDM}=-{i\over2}\;d_{l}\;\bar{l}\sigma^{\mu\nu}\gamma_5
l\;F_{{\mu\nu}}\label{edm},
\end{eqnarray}
where $\sigma_{\mu\nu}=i[\gamma_\mu,\gamma_\nu]/2$, $l$ denotes the lepton
fermion, $m_{l}$ represents the corresponding lepton mass and $F_{\mu\nu}$ is the electromagnetic field strength. $a_l$ and $d_l$ are respectively the lepton MDMs and EDMs.

To obtain the lepton MDMs and EDMs,
we use the effective Lagrangian method, the reason is that the masses of internal lines are much heavier than that of external lepton masses in the EBLMSSM. The Feynman amplitudes can be expressed by the following dimension-6 operators.
\begin{eqnarray}
&&\mathcal{O}_1^{\mp}=\frac{1}{(4\pi)^2}\bar{l}(i\mathcal{D}\!\!\!\slash)^3\omega_{\mp}l,
~~~~~~~~~~~~~~~~\mathcal{O}_2^{\mp}=\frac{eQ_f}{(4\pi)^2}\overline{(i\mathcal{D}_{\mu}l)}\gamma^{\mu}
F\cdot\sigma\omega_{\mp}l,
\nonumber\\
&&\mathcal{O}_3^{\mp}=\frac{eQ_f}{(4\pi)^2}\bar{l}F\cdot\sigma\gamma^{\mu}
\omega_{\mp}(i\mathcal{D}_{\mu}l),
~~~~~~\mathcal{O}_4^{\mp}=\frac{eQ_f}{(4\pi)^2}\bar{l}(\partial^{\mu}F_{\mu\nu})\gamma^{\nu}
\omega_{\mp}l,\nonumber\\&&
\mathcal{O}_5^{\mp}=\frac{m_l}{(4\pi)^2}\bar{l}(i\mathcal{D}\!\!\!\slash)^2\omega_{\mp}l,
~~~~~~~~~~~~~~~~\mathcal{O}_6^{\mp}=\frac{eQ_fm_l}{(4\pi)^2}\bar{l}F\cdot\sigma
\omega_{\mp}l,
\end{eqnarray}
with $\mathcal{D}_{\mu}=\partial_{\mu}+ieA_{\mu}$ and $\omega_{\mp}=(1\mp\gamma5)/2$.
Adopting on-shell condition for external lepton, only
$\mathcal{O}_{2,3,6}^{\mp}$ have contributions to lepton MDMs and EDMs. Therefore, we only study the Wilson coefficients of the operators $\mathcal{O}_{2,3,6}^{\mp}$ in the effective Lagrangian, which can be written as $C_{2,3,6}^{\mp}$. Actually, the Wilson coefficients satisfy the relations $C_{2}^{\mp}=C_{3}^{\mp*}$ and $C_{6}^{+}=C_{6}^{-*}$.

After simplifying the concerned terms in the effective Lagrangian, the lepton MDMs and EDMs are deduced as
\begin{eqnarray}
&&a_l=\frac{4eQ_fm_l^2}{(4\pi)^2}\Re(C_2^++C_2^{-*}+C_6^+),
\nonumber\\&&d_l=-\frac{2eQ_fm_l}{(4\pi)^2}\Im(C_2^++C_2^{-*}+C_6^+).
\end{eqnarray}
Here, $\Re(...)$ denotes that the lepton MDMs are proportional to the real part of effective couplings, as well as $\Im(...)$ denotes that the lepton EDMs are proportional to the imaginary part of effective couplings.
\subsection{The one-loop corrections }
\begin{figure}[t]
\centering
\includegraphics[width=11cm]{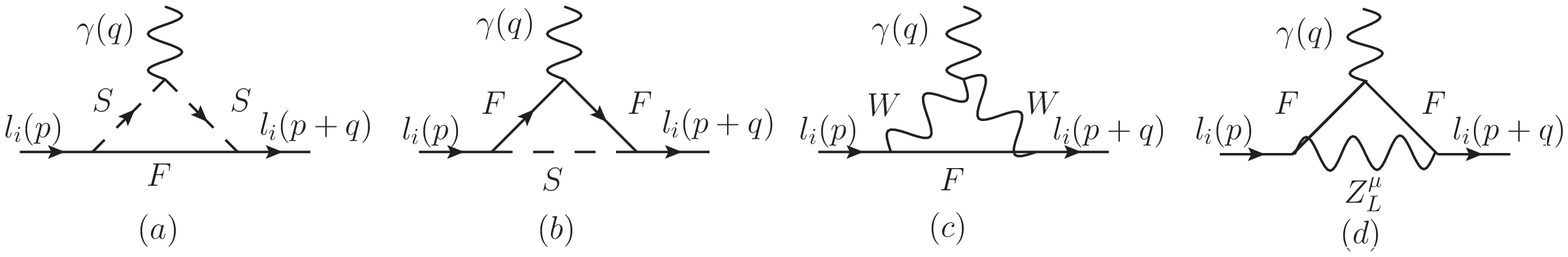}\\
\caption{The one-loop diagrams affect lepton MDMs and EDMs in the EBLMSSM.} \label{fig1}
\end{figure}

In EBLMSSM, there are new contributions to lepton MDMs and EDMs at one-loop level, which come from the triangle diagrams shown in FIG. \ref{fig1}. These new contributions come from
the tree level couplings such as neutralino-slepton, lepton neutralino-slepton, exotic slepton-$\tilde{Y}$, chargino-sneutrino, exotic lepton-$Y$, $W$-neutrino and $Z_L^{\mu}$-lepton. We have neglected the neutral Higgs-lepton and charged Higgs-neutrino
contributions to lepton MDMs and EDMs because the related Yukawa couplings are very tiny, which contain the depression factor $\frac{m_{l^I}^2}{\Lambda^2}$ and $\frac{m^2_{\nu^I}}{\Lambda^2}$($\Lambda$ representing the energy scale of new physics and $\Lambda=1$ TeV in our following calculation, $\frac{m_{\tau}^2}{\Lambda^2}\sim10^{-6}$ and $\frac{m^2_{\nu^{\tau}}}{\Lambda^2}\sim0$).

In the EBLMSSM, the lepton MDMs and EDMs corresponding to FIG. \ref{fig1}(a) are deduced as follows
\begin{eqnarray}
&&a_l(a)=
-\sum_{F=\chi^0/\chi_{NL}^0,\tilde{Y}}\sum_{S=\tilde{L},\tilde{E}}\Big[\Re[(\mathcal{S}_1)^I(\mathcal{S}_2)^{I*}]
x_S\sqrt{x_Fx_{m_{l^I}}}\;\frac{\partial^2 \mathcal{B}(x_F,x_S)}{\partial x_S^2}
\nonumber\\&&\hspace{1.4cm}+\frac{1}{3}(|(\mathcal{S}_1)^I|^2+|(\mathcal{S}_2)^I|^2)x_Sx_{m_{l^I}}
\frac{\partial\mathcal{B}_1(x_F,x_S)}{\partial x_S}\Big],
\nonumber\\&&d_l(a)=
-\sum_{F=\chi^0/\chi_{NL}^0,\tilde{Y}}\sum_{S=\tilde{L},\tilde{E}}\frac{e}{2\Lambda}\Big[\Im[(\mathcal{S}_1)^{I*}(\mathcal{S}_2)^I]
x_S\sqrt{x_F}\;\frac{\partial^2 \mathcal{B}(x_F,x_S)}{\partial x_S^2}\Big],
\end{eqnarray}
where $x_i$ denoting $\frac{m_i^2}{\Lambda^2}$ and $m_i$ representing the related masses of particles. $\mathcal{B}(x,y),\;\mathcal{B}_1(x,y)$ are the one-loop functions
 \begin{eqnarray}
\mathcal{B}(x,y)=\frac{1}{16 \pi
   ^2}\Big(\frac{x \ln x}{y-x}+\frac{y \ln
   y}{x-y}\Big),~~~
\mathcal{B}_1(x,y)=(
\frac{\partial }{\partial y}+\frac{y}{2}\frac{\partial^2 }{\partial y^2})\mathcal{B}(x,y).
\end{eqnarray}
The concrete forms of the couplings $(\mathcal{S}_1)^I,\;(\mathcal{S}_2)^{I}$ are
\begin{eqnarray}
&&(\mathcal{S}_1)^I_{\chi^0\tilde{L}}=\sum_{i=1}^6\sum_{j=1}^4\Big[\frac{e}{\sqrt{2}s_Wc_W}Z_{\tilde{L}}^{Ii}(Z_N^{1j}s_W+Z_N^{2j}c_W)+Y_l^IZ_{\tilde{L}}^{(I+3)i}Z_N^{3j}\Big],
\nonumber\\&&(\mathcal{S}_2)^I_{\chi^0\tilde{L}}=\sum_{i=1}^6\sum_{j=1}^4\Big[-\frac{\sqrt{2}e}{c_W}Z_{\tilde{L}}^{(I+3)i}Z_N^{1j*}+Y_l^IZ_{\tilde{L}}^{Ii}Z_N^{3j*}\Big];
\nonumber\\&&(\mathcal{S}_1)^I_{\chi_{NL}^0\tilde{L}}=\sqrt{2}\sum_{i=1}^6\sum_{j=1}^5g_LZ_{N_L}^{1j}Z_{\tilde{L}}^{Ii},
~~(\mathcal{S}_2)^I_{\chi_{NL}^0\tilde{L}}=-\sqrt{2}\sum_{i=1}^6\sum_{j=1}^5g_LZ_{N_L}^{1j*}Z_{\tilde{L}}^{(I+3)i};
\nonumber\\&&(\mathcal{S}_1)^I_{\tilde{Y}\tilde{E}}=\sum_{i=1}^4\lambda_4Z_{\tilde{E}}^{4i*},
~~~~(\mathcal{S}_2)^I_{\tilde{Y}\tilde{E}}=-\sum_{i=1}^4\lambda_6Z_{\tilde{E}}^{3i*}.
\end{eqnarray}

Similarly, the lepton MDMs and EDMs for FIG. \ref{fig1}(b) can be formulated as
\begin{eqnarray}
&&a_l(b)=
\sum_{F=\chi^{\pm},L'}\sum_{S=\tilde{\nu},Y}\Big[-2\Re[(\mathcal{S}_1)^I(\mathcal{S}_2)^{I*}]
\sqrt{x_Fx_{m_{l^I}}}\;\mathcal{B}_1(x_S,x_F)
\nonumber\\&&\hspace{1.0cm}+\frac{1}{3}(|(\mathcal{S}_1)^I|^2+|(\mathcal{S}_2)^I|^2)x_Fx_{m_{l^I}}
\frac{\partial\mathcal{B}_1(x_S,x_F)}{\partial x_F}\Big],
\nonumber\\&&d_l(b)=
\sum_{F=\chi^{\pm},L'}\sum_{S=\tilde{\nu},Y}\frac{e}{\Lambda}\Big[\Im[(\mathcal{S}_1)^{I*}(\mathcal{S}_2)^I]
\sqrt{x_F}\;\mathcal{B}_1(x_S,x_F)\Big],
\end{eqnarray}
where,
\begin{eqnarray}
&&(\mathcal{S}_1)^I_{\chi^{\pm}\tilde{\nu}}=-\sum_{i=1}^6\sum_{j=1}^2\big[\frac{e}{s_w}Z_+^{1j}Z_{\tilde{\nu}}^{Ii*}+Y_{\nu}^{Ii}Z_+^{2j}Z_{\tilde{\nu}}^{(I+3)i*}\big],
~(\mathcal{S}_2)^I_{\chi^{\pm}\tilde{\nu}}=-\sum_{i=1}^6\sum_{j=1}^2{Y_l^{I}Z_-^{2j*}Z_{\tilde{\nu}}^{Ii*}};
\nonumber\\&&(\mathcal{S}_1)^I_{L'Y}=-\sum_{i=1}^2\sum_{j=1}^2\lambda_6U_L^{2i}Z_Y^{2j*},
~~(\mathcal{S}_2)^I_{L'Y}=-\sum_{i=1}^2\sum_{j=1}^2\lambda_4W_L^{1i}Z_Y^{1j*}.
\end{eqnarray}

Then the lepton MDMs for FIG. \ref{fig1}(c) and FIG. \ref{fig1}(d) are given out
\begin{eqnarray}
&&a_l(c)=\Big[2|(\mathcal{S}_1)^I|^2x_{m_{l^I}}(2\mathcal{B}_1(x_{\nu},x_W)
+\frac{x_W}{3}\frac{\partial\mathcal{B}_1(x_{\nu},x_W)}{\partial x_W})\Big],
(\mathcal{S}_1)^I_{\nu W}=\frac{-e}{\sqrt{2}s_W}\sum_{i=1}^6Z_{N_i}^{Ii*}.
\end{eqnarray}
\begin{eqnarray}
&&a_l(d)=
g_L^2\Big[\frac{2}{3}x_{m_{l^I}}^2\frac{\partial\mathcal{B}_1(x_{Z_L^{\mu}},x_{m_{l^I}})}{\partial x_{m_{l^I}}}-8x_{m_{l^I}}(\frac{\partial\mathcal{B}(x_{Z_L^{\mu}},x_{m_{l^I}})}{\partial x_{m_{l^I}}}+\mathcal{B}_1(x_{Z_L^{\mu}},x_{m_{l^I}}))\Big].
\end{eqnarray}
Here, the one-loop contributions to lepton EDMs from FIG. \ref{fig1}(c) and FIG. \ref{fig1}(d) are zero. In our latter numerical calculations, the effects from $W$-neutrino can be ignored due to the tiny neutrino masses.

Above all, the one-loop corrections to lepton MDMs and EDMs can be expressed as:
\begin{eqnarray}
&&\Delta a_l^{one-loop}=a_l(a)+a_l(b)+a_l(d),~~d_l^{one-loop}=d_l(a)+d_l(b).
\label{oneloop}
\end{eqnarray}
\subsection{The two-loop Barr-Zee and Rainbow type corrections}
\begin{figure}[t]
\centering
\includegraphics[width=9cm]{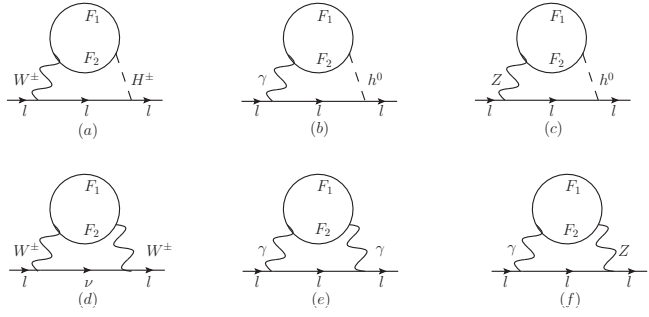}\\
\caption{The two-loop Barr-Zee and Rainbow type diagrams affect lepton MDMs and EDMs in the EBLMSSM.} \label{fig2}
\end{figure}
In this section, we discuss the contributions of two-loop Barr-Zee and Rainbow type diagrams to lepton MDMs and EDMs. The two-loop contributions are suppressed by the heavy scalar particles, such as sleptons, exotic sleptons, squarks, sneutrinos and exotic sneutrinos. So according to the decoupling theorem, we ignore these diagrams in the calculation below\cite{deltaau3}. In the EBLMSSM, we only consider the two-loop contributions that a closed fermion loop is attached to the virtual gauge bosons or the Higgs fields. Corresponding, the two-loop Barr-Zee and Rainbow type diagrams that possess a major contribution to lepton MDMs and EDMs are given out in FIG. \ref{fig2}.

First, we consider the corrections from FIG. \ref{fig2}(a). According to Ref.\cite{twoaldlreduce}, we give the analytical expressions of lepton MDMs and EDMs under the assumption $m_F=m_{F_1}=m_{F_2}\gg m_W$, which can be simplified as:
\begin{eqnarray}
&&a_l^{WH}=\frac{G_F m_l m_W^2s_W}{128e\pi^4}\sum_{F_1=\chi^{\pm},L'}\sum_{F_2=\chi^0,N'}\frac{H_{\bar{l}H\nu}^L}{ m_F}\Big\{\Big[\frac{21}{4}-\frac{5}{18}Q_{F_1}+(3+\frac{Q_{F_1}}{3})
(\ln{m_{F_1}^2}\nonumber\\
&&\qquad\quad\hspace{-0.8cm}-\varrho_{1,1}(m_W^2,m_{H^\pm}^2))\Big]\Re(H_{HF_1F_2}^LH_{WF_1F_2}^L+H_{HF_1F_2}^RH_{WF_1F_2}^R)
+\Big[\frac{19-20Q_{F_1}}{9}\nonumber\\
&&\qquad\quad\hspace{-0.8cm}+\frac{2-4Q_{F_1}}{3}
(\ln{m_{F_1}^2}-\varrho_{1,1}(m_W^2,m_{H^\pm}^2))\Big]\Re(H_{HF_1F_2}^LH_{WF_1F_2}^R+H_{HF_1F_2}^RH_{WF_1F_2}^L)
\nonumber\\
&&\qquad\quad\hspace{-0.8cm}+\Big[\hspace{-0.1cm}-\hspace{-0.1cm}\frac{16}{9}\hspace{-0.1cm}-\hspace{-0.1cm}\frac{2\hspace{-0.1cm}+\hspace{-0.1cm}6Q_{F_1}}{3}
(\ln{m_{F_1}^2}\hspace{-0.1cm}-\hspace{-0.1cm}\varrho_{1,1}(m_W^2,m_{H^\pm}^2))\Big]\Re(H_{HF_1F_2}^LH_{WF_1F_2}^L\hspace{-0.1cm}-\hspace{-0.1cm}H_{HF_1F_2}^RH_{WF_1F_2}^R)
\nonumber\\
&&\qquad\quad\hspace{-0.8cm}+\Big[\hspace{-0.1cm}-\hspace{-0.1cm}\frac{2Q_{F_1}}{9}\hspace{-0.1cm}-\hspace{-0.1cm}\frac{6\hspace{-0.1cm}-\hspace{-0.1cm}2Q_{F_1}}{3}
(\ln{m_{F_1}^2}\hspace{-0.1cm}-\hspace{-0.1cm}\varrho_{1,1}(m_W^2,m_{H^\pm}^2))\Big]\Re(H_{HF_1F_2}^LH_{WF_1F_2}^R\hspace{-0.1cm}-\hspace{-0.1cm}H_{HF_1F_2}^RH_{WF_1F_2}^L)\Big\},
\nonumber\\&&d_l^{WH}=\frac{G_Fm_W^2s_W}{256\pi^4}\sum_{F_1=\chi^{\pm},L'}\sum_{F_2=\chi^0,N'}\frac{H_{\bar{l}H\nu}^L}{ m_F}\Big\{\Big[\frac{21}{4}-\frac{5}{18}Q_{F_1}+(3+\frac{Q_{F_1}}{3})
(\ln{m_{F_1}^2}\nonumber\\
&&\qquad\quad\hspace{-0.8cm}-\varrho_{1,1}(m_W^2,m_{H^\pm}^2))\Big]\Im(H_{HF_1F_2}^LH_{WF_1F_2}^L+H_{HF_1F_2}^RH_{WF_1F_2}^R)
+\Big[\frac{19-20Q_{F_1}}{9}\nonumber\\
&&\qquad\quad\hspace{-0.8cm}+\frac{2-4Q_{F_1}}{3}
(\ln{m_{F_1}^2}-\varrho_{1,1}(m_W^2,m_{H^\pm}^2))\Big]\Im(H_{HF_1F_2}^LH_{WF_1F_2}^R+H_{HF_1F_2}^RH_{WF_1F_2}^L)
\nonumber\\
&&\qquad\quad\hspace{-0.8cm}+\Big[\hspace{-0.1cm}-\hspace{-0.1cm}\frac{16}{9}\hspace{-0.1cm}-\hspace{-0.1cm}\frac{2\hspace{-0.1cm}+\hspace{-0.1cm}6Q_{F_1}}{3}
(\ln{m_{F_1}^2}\hspace{-0.1cm}-\hspace{-0.1cm}\varrho_{1,1}(m_W^2,m_{H^\pm}^2))\Big]\Im(H_{HF_1F_2}^LH_{WF_1F_2}^L\hspace{-0.1cm}-\hspace{-0.1cm}H_{HF_1F_2}^RH_{WF_1F_2}^R)
\nonumber\\
&&\qquad\quad\hspace{-0.8cm}+\Big[\hspace{-0.1cm}-\hspace{-0.1cm}\frac{2Q_{F_1}}{9}\hspace{-0.1cm}-\hspace{-0.1cm}\frac{6\hspace{-0.1cm}-\hspace{-0.1cm}2Q_{F_1}}{3}
(\ln{m_{F_1}^2}\hspace{-0.1cm}-\hspace{-0.1cm}\varrho_{1,1}(m_W^2,m_{H^\pm}^2))\hspace{-0.05cm}\Big]\hspace{-0.1cm}\Im(H_{HF_1F_2}^LH_{WF_1F_2}^R\hspace{-0.1cm}-\hspace{-0.1cm}H_{HF_1F_2}^RH_{WF_1F_2}^L\hspace{-0.1cm})\hspace{-0.1cm}\Big\},
\end{eqnarray}
where $\varrho_{1,1}(x,y)=\frac{x\ln x-y\ln y}{x-y}$, $H_{HF_1F_2}^{L,R}$ and $H_{WF_1F_2}^{L,R}$ represent the coupling coefficients of the corresponding vertices. The concrete contributions of other two-loop Barr-Zee and Rainbow type diagrams are compiled into the Appendix A. Then the two-loop contributions to lepton MDMs and EDMs can be summarized as
\begin{eqnarray}
&&\Delta a_l^{two-loop}=\Delta a_l^{one-loop}+a_l^{WH}+a_l^{\gamma h}+a_l^{Z h}+a_l^{WW}+a_l^{\gamma\gamma}+a_l^{\gamma Z},
\nonumber\\&&d_l^{two-loop}=d_l^{one-loop}+d_l^{WH}+d_l^{\gamma h}+d_l^{Z h}+d_l^{WW}.
\label{twoloop}
\end{eqnarray}
\section{The numerical results}
The numerical results are discussed in this section. The lightest CP-even Higgs mass is considered as an input parameter, which is around $m_{h^0}=125.18$ GeV. We consider the constrains from the processes $h^0\rightarrow \gamma\gamma$, $h^0\rightarrow VV, V=(Z,W)$ discussed in our previous work\cite{EBL1}, which confines the parameter space of the EBLMSSM. As well as, the constrains from the charged lepton flavor violation (CLFV) processes in the EBLMSSM should not be ignored\cite{EBL2}, such as processes $l_j\rightarrow l_i \gamma$, $\mu-e$ conversion in nuclei, the $\tau$ decays and $h^0\rightarrow l_i l_j$.

The used parameters in the EBLMSSM are given out as follows:
\begin{eqnarray}
&&Y_{\nu_4} = Y_{\nu_5} = 0.8,
~m_{\tilde{\nu}_4} = m_{\tilde{\nu}_5}=A_{\nu_4} = A_{\nu_5}=1{\rm TeV},v_{Nlt}=v_{lt}=3{\rm TeV},\nonumber\\&&L_4 = 1.5,
~\tan\beta_L= 2,~(\lambda_{Nc})_{ii} = 1,
~(m_{\tilde{\nu}})_{ii}= 1{\rm TeV},~(A'_l)_{ii}=0.3{\rm TeV},
\nonumber\\&&
(m_{\tilde{L}}^2)_{ij}=1000{\rm GeV}^2,~~(A_{N})_{ii} =(A_{Nc})_{ii} = 0.5{\rm TeV},~i,j=1.2.3,~i\neq j. \label{canshu}
\end{eqnarray}

The following assumptions are adopted to simplify the numerical discussion:
\begin{eqnarray}
&&(m_{\tilde{L}}^2)_{ii}=S_m^2,~A_{LL}=A_{LE} = A_{LN} = A_E,~A_{e_4} =A_{e_5} = A_{\tilde{E}},~\lambda_L =\lambda_E=\lambda_{N_L} = L_l,\nonumber\\&&(A_l)_{ii}=A_l,~m_{\tilde{L}_4} = m_{\tilde{L}_5} =m_{\tilde{e}_4} = m_{\tilde{e}_5} = M_{\tilde{E}},(\lambda_4^2)_{II} =(\lambda_6^2)_{II} = (Lm^2)_{II},I=1,2,3,\nonumber\\&&
\mu=|\mu|e^{i\theta_{\mu}},~~m_1=|m_1|e^{i\theta_{1}},~~m_2=|m_2|e^{i\theta_{2}},~~\mu_L=|\mu_L|e^{i\theta_{\mu_L}},\nonumber\\&&
M_L=|M_L|e^{i\theta_{M_L}},\mu_{NL}=|\mu_{NL}|e^{i\theta_{NL}},\mu_Y=|\mu_Y|e^{i\theta_{\mu_Y}},B_Y=|B_Y|e^{i\theta_{B_Y}}.
\end{eqnarray}
We take $\sqrt{(Lm^2)_{11}}=L_S$ and $\sqrt{(Lm^2)_{22}}=\sqrt{(Lm^2)_{33}}=L_s$. $\theta_{\mu}$, $\theta_{1}$, $\theta_{2}$, $\theta_{M_L}$, $\theta_{\mu_L}$, $\theta_{NL}$, $\theta_{\mu_Y}$, $\theta_{B_Y}$ represent the CP-violating phases corresponding to parameters $\mu$, $m_1$, $m_2$, $M_L$, $\mu_L$, $\mu_{NL}$, $\mu_Y$, $B_Y$.
\subsection{The electron MDM and EDM}
In this section, we study the two-loop contributions to the electron MDM and EDM. We consider $2.4\sigma$ experimental error to electron MDM, which is constrained as $-17.44\times 10^{-13}<\Delta a_{e}<-0.14\times 10^{-13}$. The present experimental upper bound of electron EDM is $|d_e|<8.7\times10^{-29}$ e.cm, which is the most strict one for new physics.

\begin{figure}
\centering
\begin{minipage}[c]{0.48\textwidth}
\includegraphics[width=6cm]{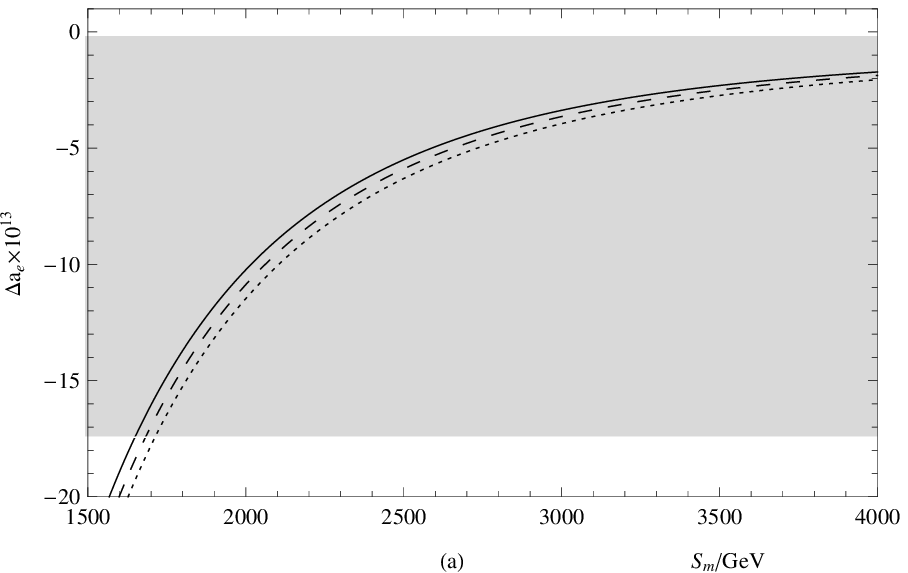}
\end{minipage}%
\begin{minipage}[c]{0.45\textwidth}
\includegraphics[width=6cm]{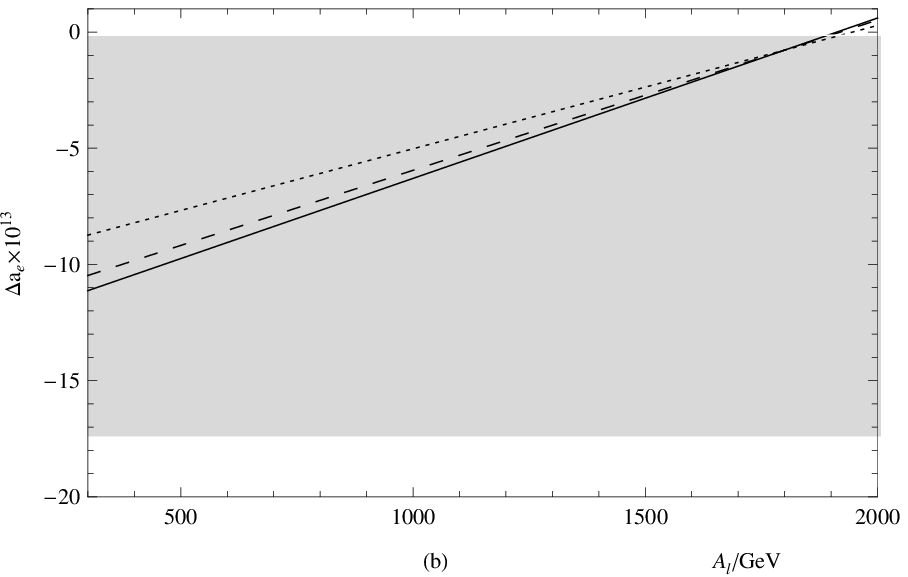}
\end{minipage}
\caption[]{The electron MDM $\Delta a_{e}$ varying with the parameter $S_m$ ($A_l$) are plotted by dotted line, dashed line and solid line respectively in FIG. \ref{ale}(a) (FIG. \ref{ale}(b)) when $g_L=0.05,0.10,0.16$ ( $m_1=0.5,0.8,1.1$ TeV). The gray area denotes the experimental 2.4$\sigma$ interval.}\label{ale}
\end{figure}

First, we consider the two-loop contributions to electron MDM $\Delta a_e$ versus parameter $S_m$ in FIG. \ref{ale}(a), where $\mu=0.7$ TeV, $m_1=m_2=A_l=\mu_L=1.0$ TeV and $M_L=1.5$ TeV. $g_L=0.05(0.10,0.16)$ corresponds to dotted (dashed, solid) line. These three lines almost overlap, which demonstrates that parameter $g_L$ has small contributions to the numerical results. Besides, when $S_m$  is greater than 1.6 TeV, $\Delta a_e$ increases with the increase of $S_m$ within a reasonable deviation range, and gradually approaches 0. This indicates that $\Delta a_e$ is decoupling with the increase of $S_m$. In addition, the absolute values of two-loop diagrams' corrections to the one-loop predictions are around $0.01\%\sim0.2\%$ with the enlarging $S_m$.

Then, the strong impacts of parameter $A_l$ on the two-loop contributions to electron MDM is further illustrated by FIG. \ref{ale}(b), In order to obtain the suitable results, we assume that $g_L=0.10$ and $S_m=\sqrt{6}$ TeV. Furthermore, the results of electron MDM in the dotted line, dashed line and solid line correspond to $m_1=0.5,0.8,1.1$ TeV respectively. The figure shows that $\Delta a_e$ decreases slightly with increasing $m_1$. Additionally, the results of these three lines, agreeing well with the deviation between SM prediction and experimental results, all have obvious increase when $A_l$ increases from 0.3 to 1.8 TeV. Meanwhile, the absolute values of two-loop diagrams' corrections to one-loop contributions vary from $0.03\%$ to $0.4\%$. So parameter $A_l$ affects the electron MDM remarkably.

Supposing CP-violating phases $\theta_{\mu}=\theta_{1}=\theta_{2}=\theta_{NL}=\theta_{M_L}=0$, we study the two-loop contributions to the electron EDM. Many literatures\cite{cancellation} have studied the cancellation scenario for electron EDM in order to induce it below the experimental upper bounds. In the discussion below, we look for the contributions of new CP-violating phases to electron EDM under the premise of satisfying the cancellation mechanism.
\begin{figure}
\centering
\begin{minipage}[c]{0.33\textwidth}
\includegraphics[width=5cm]{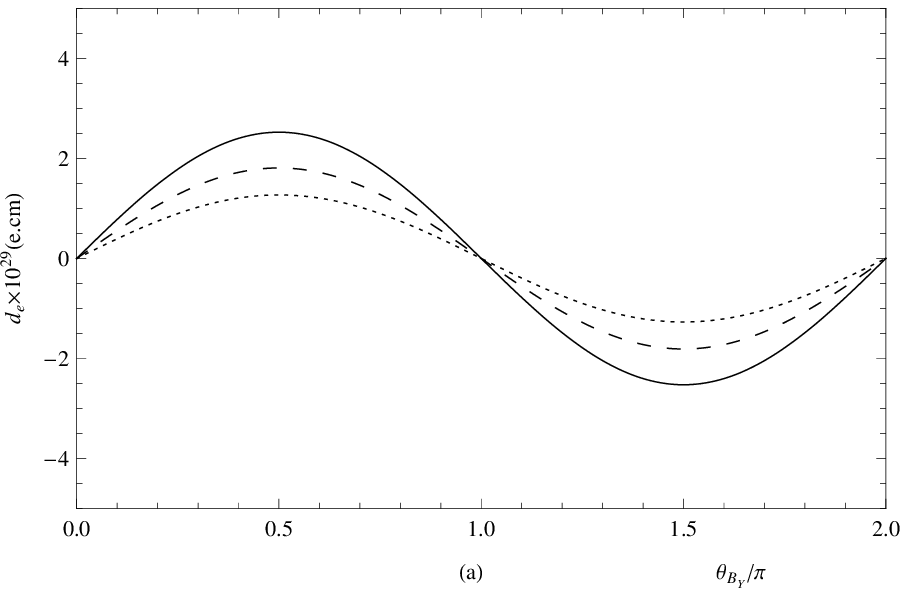}
\end{minipage}%
\begin{minipage}[c]{0.33\textwidth}
\includegraphics[width=5cm]{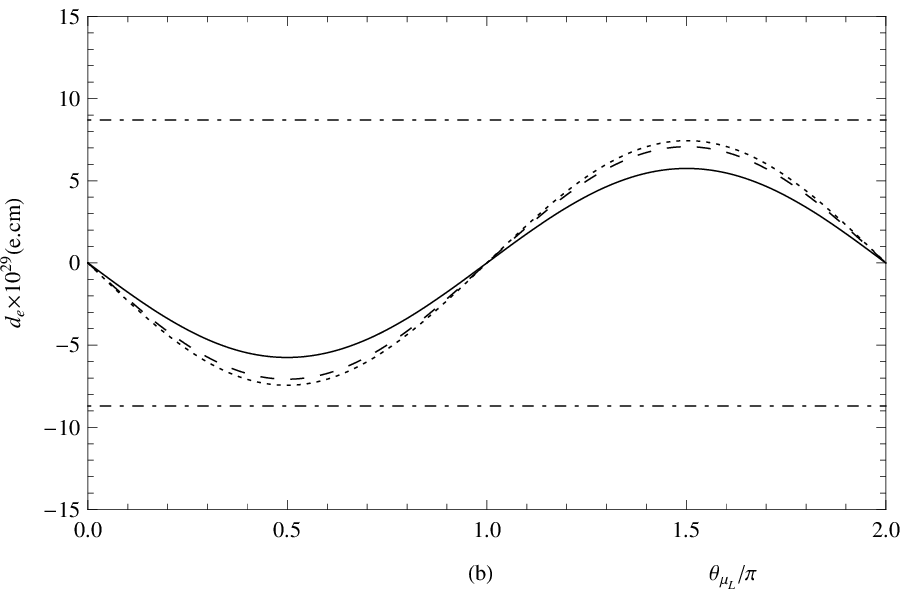}
\end{minipage}
\begin{minipage}[c]{0.33\textwidth}
\includegraphics[width=5cm]{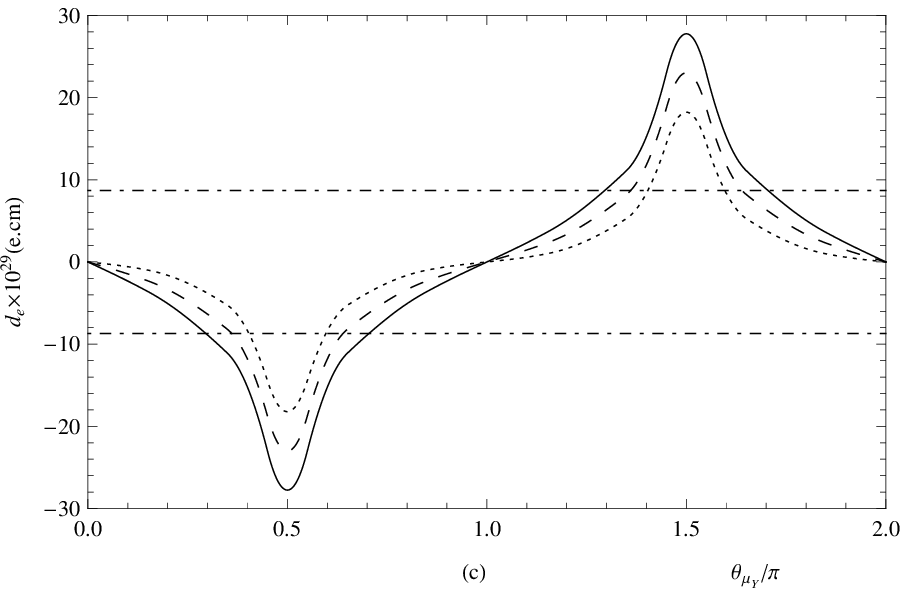}
\end{minipage}
\caption[]{With $|B_Y|=0.8,1.0,1.2$ TeV ($|\mu_L|=1.0,2.0,3.0$ TeV, $Y_{e_5}=0.8,1.1,1.4$), the electron EDM $d_{e}$ varying with the CP-violating phase $\theta_{B_Y}$ ($\theta_{\mu_L}$, $\theta_{\mu_Y}$) are plotted by dotted line, dashed line and solid line respectively in FIG. \ref{dle}(a)(FIG. \ref{dle}(b), FIG. \ref{dle}(c)). The dotdashed lines correspond to $|d_e|=8.7\times10^{-29}$ e.cm.}\label{dle}
\end{figure}

With $L_S=0.001$, the two-loop contributions to electron EDM $d_{e}$ varying with the CP-violating phases $\theta_{B_Y}$, $\theta_{\mu_L}$ and $\theta_{\mu_Y}$ are respectively plotted in FIG. \ref{dle} (a), (b) and (c). Accordingly, the dotted line, dashed line and solid line respectively correspond to $|B_Y|=0.8,1.0,1.2$ TeV ($|\mu_L|=1.0,2.0,3.0$ TeV, $Y_{e_5}=0.8,1.1,1.4$). It is easy to see that $d_{e}$ shows a sinusoidal trend as $\theta_{B_Y}$ changes from 0 to $2\pi$, while possesses a cosine trend as $\theta_{\mu_L}$ varies from 0 to $2\pi$. Not only that, the absolute values of $d_{e}$ increases with the increase of $|B_Y|$ but decreases with the increase of $|\mu_L|$. Additionally, $d_{e}$ varying with $\theta_{B_Y}$ and $\theta_{\mu_L}$ both meet the experimental upper limit and have maximums when $\theta_{B_Y}=\theta_{\mu_L}=\pm0.5\pi$. From FIG.\ref{dle} (c), we find that $d_{e}$ have smart change when $\theta_{\mu_Y}$ varies in the region $0\sim 2\pi$. Especially, the $d_{e}$ exceeds the experimental upper bound when $\theta_{\mu_Y}$ are around $\pm(0.3\pi\sim0.7\pi)$. Therefore, $\theta_{\mu_Y}$ influences the electron EDM $d_{e}$ very obviously. Furthermore, the smaller $Y_{e_5}$, the larger parameter range for $\theta_{\mu_Y}$ that satisfies the experimental upper bound.
\subsection{The muon EDM and MDM}
In this section, we discuss the two-loop corrections to muon MDM and EDM. The $3.7\sigma$ experimental deviation between SM prediction and experimental result in Eq.(\ref{deltaau}) is considered for muon MDM, which is $0.39\times10^{-10}\sim  54.41\times10^{-10}$. Moreover, the present experimental upper bound of muon EDM is $|d_{\mu}|<1.9\times10^{-19}$ e.cm. Here, we suppose $\tan\beta_{NL}=2$, $A_{\tilde{E}}=1.0$ TeV, $\mu_Y=2$ TeV, $B_Y=1.0$ TeV and $L_l=L_s=1$.

\begin{figure}
\centering
\begin{minipage}[c]{0.48\textwidth}
\includegraphics[width=6cm]{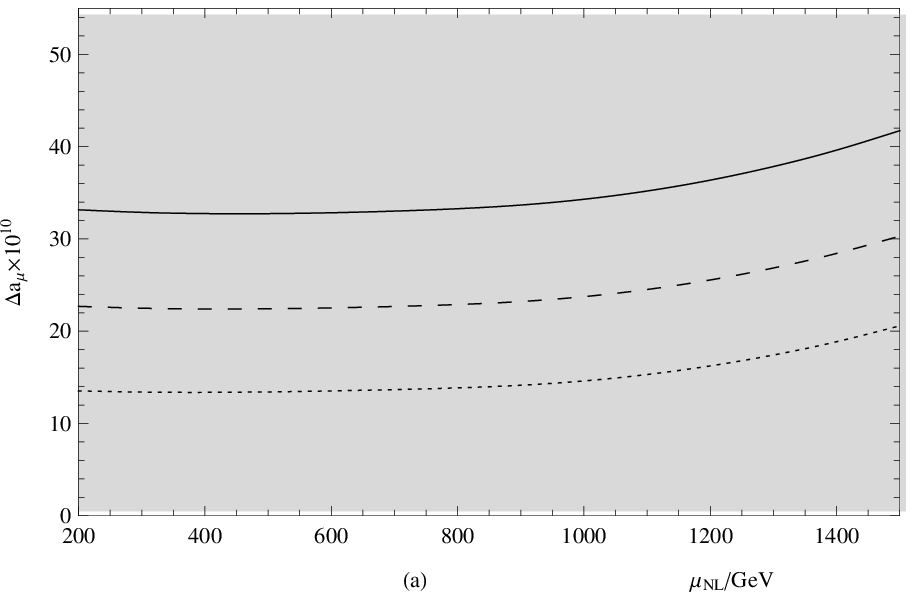}
\end{minipage}%
\begin{minipage}[c]{0.45\textwidth}
\includegraphics[width=6cm]{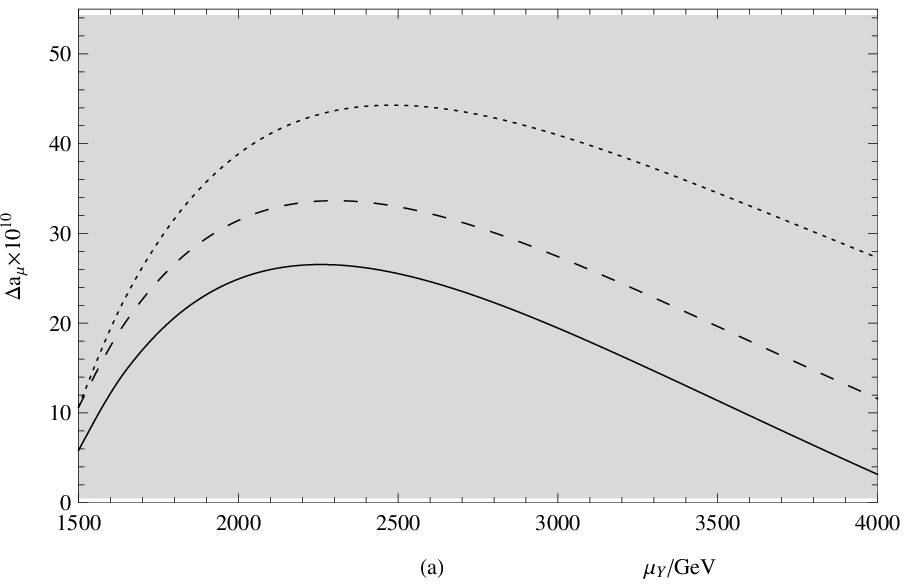}
\end{minipage}
\caption[]{The muon MDM $\Delta a_{\mu}$ varying with the parameter $\mu_{NL}$ ($\mu_Y$) are plotted by dotted line, dashed line and solid line respectively in FIG. \ref{alu}(a) (FIG. \ref{alu}(b)) when $A_E=0.8,1.0,1.2$ TeV ($\tan\beta=6,16,26$).  The gray area denotes the experimental 3.7$\sigma$ interval.}\label{alu}
\end{figure}

With all the CP-violating phases being taken zero, we study the two-loop corrections to muon MDM $\Delta a_{\mu}$ as follows. Choosing $M_{\tilde{E}}=1.8$ TeV, the muon MDM versus parameter $\mu_{NL}$ and $\mu_Y$ are respectively researched in FIG. \ref{alu} (a) and FIG. \ref{alu} (b). Here, the dotted line, dashed line and solid line respectively represent $A_E=0.8,1.0,1.2$ TeV ($\tan\beta=6,16,26$). Through the diagrams we can see that our numerical results are all within the experimental deviation range. Besides, $\Delta a_{\mu}$ increases with the enlarging $\mu_{NL}$ and $A_E$, while decreases with the enlarging $\tan\beta$. As $\mu_Y$ is around 2.2 TeV, we can obtain the largest $\Delta a_{\mu}$. Through calculation, we find that the two-loop diagrams' effects can correct the one-loop predictions in FIG. \ref{alu}(a) up to $1.2\%$, the larger $A_E$, the smaller correction it is. Furthermore, the two-loop diagrams' correction in FIG. \ref{alu}(b) can also reach up to $1.4\%$.

\begin{figure}
\centering
\begin{minipage}[c]{0.32\textwidth}
\includegraphics[width=5cm]{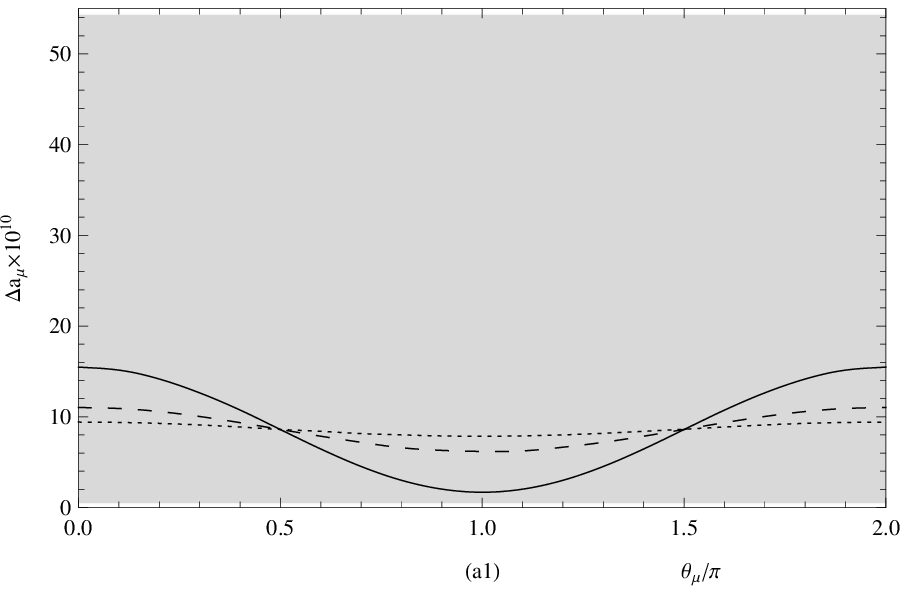}
\end{minipage}
\begin{minipage}[c]{0.32\textwidth}
\includegraphics[width=5cm]{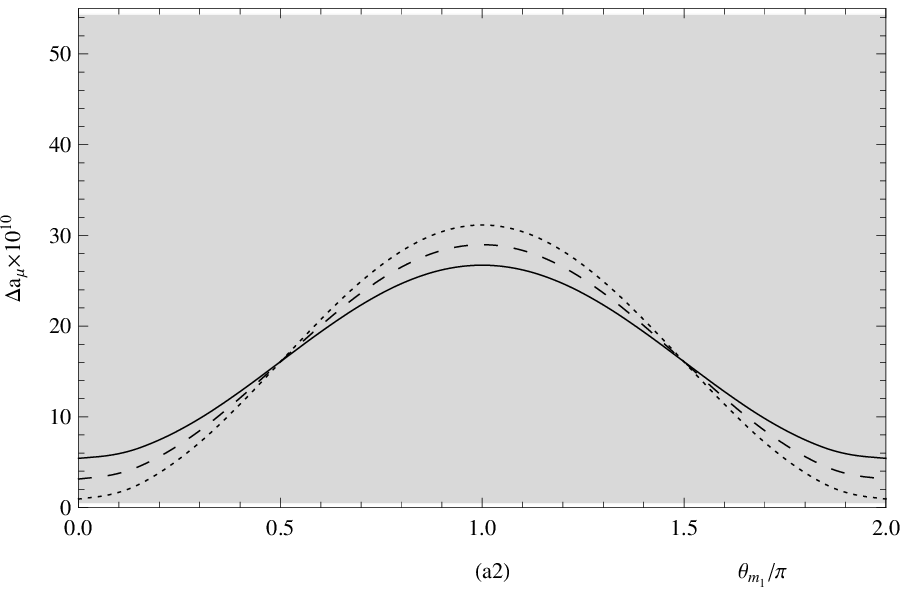}
\end{minipage}
\begin{minipage}[c]{0.32\textwidth}
\includegraphics[width=5cm]{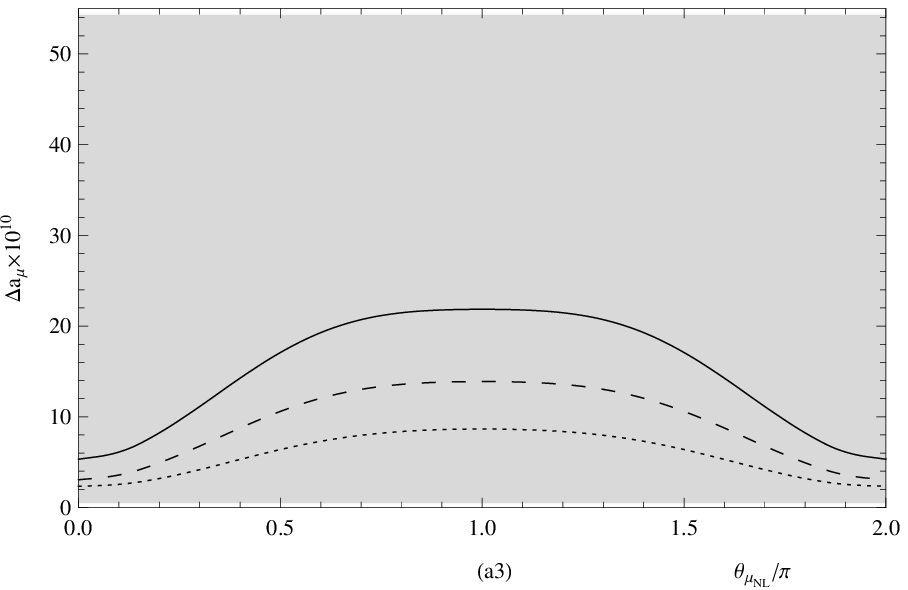}
\end{minipage}\\
\begin{minipage}[c]{0.32\textwidth}
\includegraphics[width=5cm]{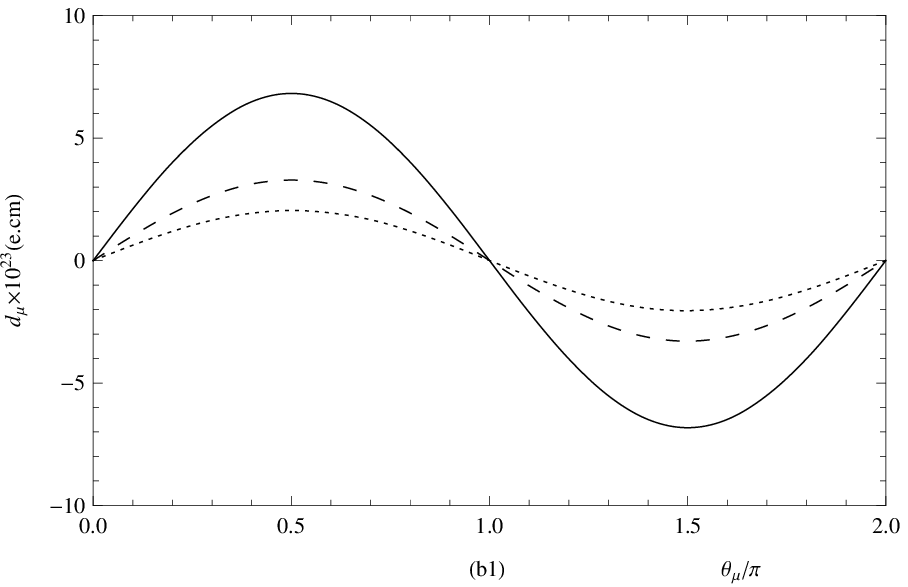}
\end{minipage}
\begin{minipage}[c]{0.32\textwidth}
\includegraphics[width=5cm]{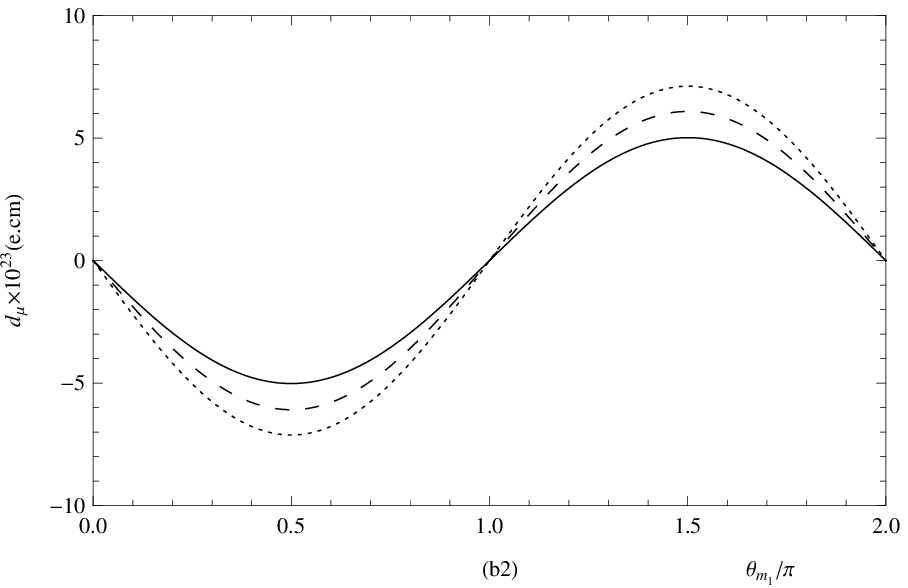}
\end{minipage}
\begin{minipage}[c]{0.32\textwidth}
\includegraphics[width=5cm]{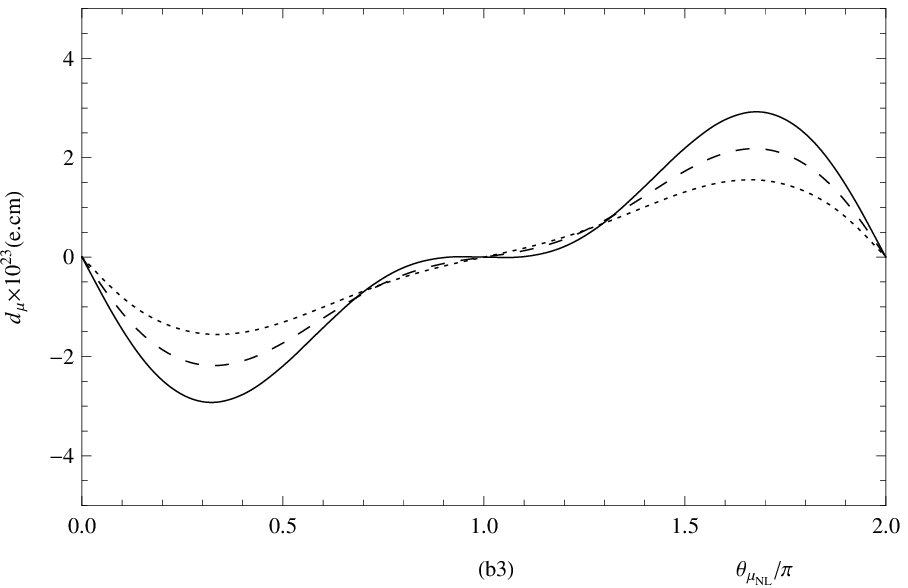}
\end{minipage}%
\caption[]{The muon MDM $\Delta a_{\mu}$ and EDM $d_{\mu}$ vary with the CP-violating phase $\theta_{\mu}$ ($\theta_{1}$, $\theta_{NL}$) in FIG. \ref{aldlu}(a1) and (b1)(FIG. \ref{aldlu}(a2) and (b2), FIG. \ref{aldlu}(a3) and (b3)), where the dotted line, dashed line and solid line stand for parameter $|\mu|=1.2,2.1,3.0$ TeV ($|m_1|=0.7,1.7,2.7$ TeV, $|\mu_{NL}|=0.8,1.1,1.4$ TeV) respectively. The gray area denotes the experimental 3.7$\sigma$ interval. }\label{aldlu}
\end{figure}
Next we will discuss the effects of the CP-violating phases $\theta_{\mu}$, $\theta_{1}$ and $\theta_{NL}$ on two-loop corrections to the muon MDM $\Delta a_{\mu}$ and muon EDM $d_{\mu}$. At first, we study $\Delta a_{\mu}$ and $d_{\mu}$ varying with the CP-violating phase $\theta_{\mu}$, where the dotted line, dashed line and solid line stand for parameter $|\mu|=1.2,2.1,3.0$ TeV respectively. In the parameter space we took, the $\Delta a_{\mu}$ can account for the deviation between SM prediction and experimental data well, which can be summarized in FIG. \ref{aldlu}(a1). Meanwhile, the two-loop diagrams' corrections to one-loop results are in the region of $0.7\%\sim7\%$. As for the muon EDM $d_{\mu}$, FIG. \ref{aldlu}(b1) tells us that the $|d_{\mu}|$ can reach $5\times10^{-23}$ e.cm at the largest CP-violating $\theta_{\mu}=\pm0.5\pi$ with $|\mu|=3.0$ TeV. In addition, the larger $|\mu|$, the larger $|d_{\mu}|$ it is.

Then, taking $M_{\tilde{E}}=1.9$ TeV, the two-loop corrections to $\Delta a_{\mu}$ and $d_{\mu}$ versus $\theta_{1}$ are researched in FIG. \ref{aldlu}(a2) and FIG. \ref{aldlu}(b2) accordingly. It is clear to see that the results of muon MDM in FIG. \ref{aldlu}(a2) are all lie in the range of $3.7\sigma$, which evidence that the parameters used in this part are all feasible. And the larger $|m_1|$, the smaller range of $\Delta a_{\mu}$ is obtained. The corrections from the two-loop diagrams should not be underestimated, the reason is that the contributions from the two-loop diagrams account for $1\%\sim10\%$ of the one-loop corrections. However, not only the $\Delta a_{\mu}$ in FIG. \ref{aldlu}(a2) rarely changes with the variational parameter $|m_1|$, but also FIG. \ref{aldlu}(b2) has very small fluctuations with $|m_1|$. So $|m_1|$ affects muon EDM and MDM weakly. When $\theta_{1}$ varies from 0 to $2\pi$, $d_{\mu}$ satisfies the experimental upper limit and acquires the maximum at $\theta_{1}=\pm0.5\pi$.

As a new introduced parameter in the EBLMSSM, $\mu_{NL}$ presents in mass matrices of exotic slepton and lepton neutralino, and influences the muon MDM and EDM through exotic slepton-$\tilde{Y}$ and slepton-exotic neutralino diagrams. So the relevant parameter $|\mu_{NL}|$ and CP-violating phase $\theta_{NL}$ are worth studying. We plot the muon MDM $\Delta a_{\mu}$ and EDM $d_{\mu}$ in FIG. \ref{aldlu}(a3) and (b3) with CP-violating phase $\theta_{NL}$, where the dotted (dashed, solid) line corresponds to $|\mu_{NL}|=0.8(1.1,1.4)$ TeV. FIG. \ref{aldlu}(a3) gives out that the parameters taken here satisfy the 3.7$\sigma$ deviation. $\Delta a_{\mu}$ increases with the enlarging $|\mu_{NL}|$ even obtains the maximum as $\theta_{NL}=\pi$. As CP-violating phase $\theta_{NL}=\pm0.4\pi$, we can obtain the biggest absolute values of muon EDM, which are under the present experimental upper bounds.
\subsection{The tau EDM}
\begin{figure}
\centering
\begin{minipage}[c]{0.45\textwidth}
\includegraphics[width=6cm]{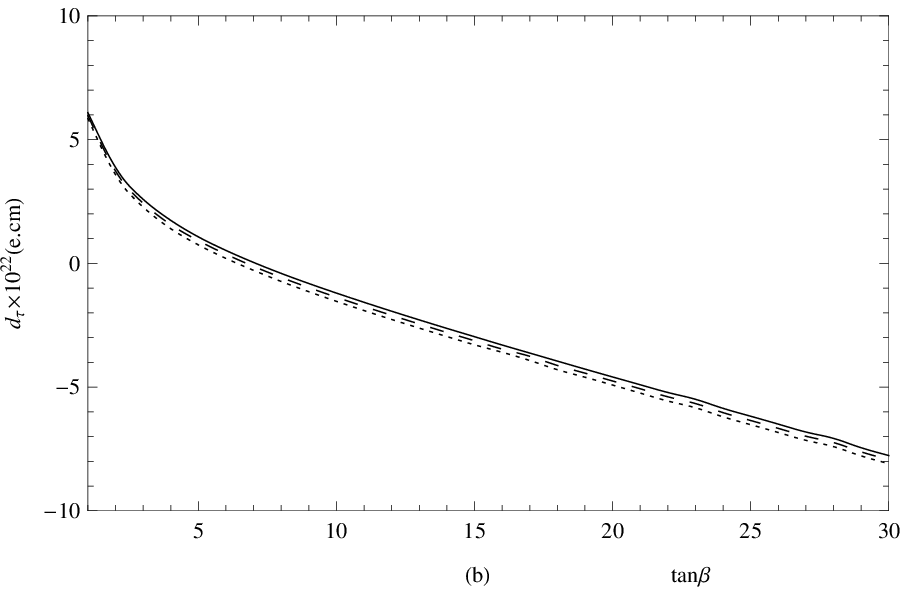}
\end{minipage}%
\begin{minipage}[c]{0.45\textwidth}
\includegraphics[width=6cm]{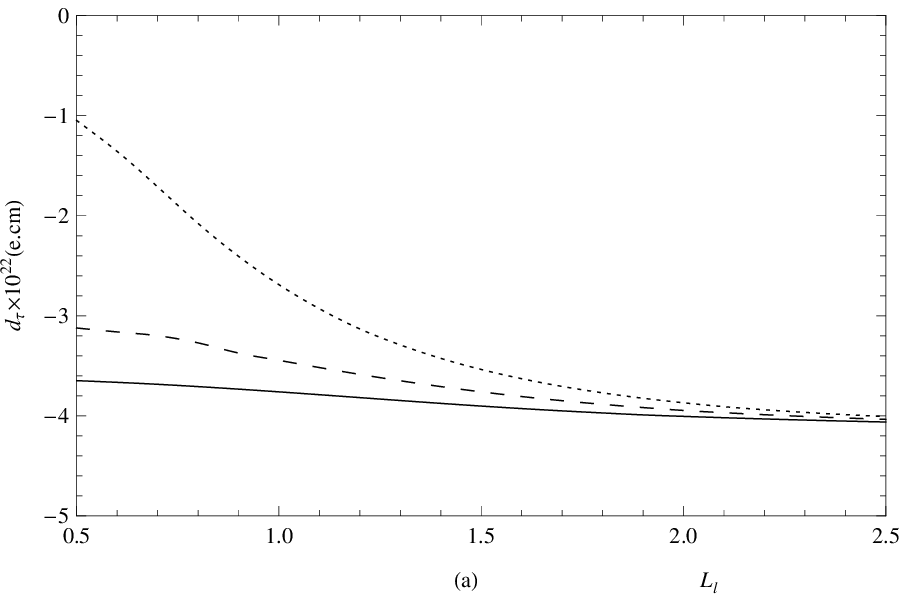}
\end{minipage}
\caption[]{With $\theta_{\mu}=0.5\pi$ and $Y_{e_4}=0.8,1.0,1.2$ ($M_{\tilde{E}}=1.0,1.5,2.0$ TeV), the two-loop contributions to tau EDM $d_{\tau}$ varying with parameter $\tan\beta$ ($L_l$) are plotted by dotted line, dashed line and solid line respectively in FIG. \ref{dltau}(a)(FIG. \ref{dltau}(b)).}\label{dltau}
\end{figure}

The present experimental upper bound of tau EDM is $|d_{\tau}|<1\times10^{-17}$ e.cm, which is the largest one among the lepton EDMs. We study the tau EDM in this subsection to obtain a more meaning result. Parameters $m_2=0.5$ TeV and $S_m=2.0$ TeV are taken into account in this part. In FIG. \ref{dltau}(a) and (b), the two-loop contributions to tau EDM $d_{\tau}$ versus $\tan\beta$ and $L_l$ are studied when CP-violating phase $\theta_{\mu}$ is $0.5\pi$. Accordingly, the numerical results are plotted with $Y_{e_4}=0.8,1.0,1.2$ ($M_{\tilde{E}}=1.0,1.5,2.0$ TeV), which are expressed by dotted line, dashed line and solid line respectively. In FIG. \ref{dltau}(a), the three lines possess the similar variation trend and the corresponding results at each point are almost the same. So the effects from parameter $Y_{e_4}$ are very weak. Other than this, the results of tau EDM have a very conspicuous decline with the $\tan\beta$ changing from 1 to 40. So parameter $\tan\beta$ plays a very important role to the tau EDM in the EBLMSSM. FIG. \ref{dltau}(b) shows that the values of tau EDM all drop smoothly as $L_l$ increases, and large $M_{\tilde{E}}$ has depressing effects on $d_{\tau}$.
\section{discussion and conclusion}
Applying the effective Lagrangian method, we study the two-loop corrections to the lepton MDMs and EDMs in the CP-violating EBLMSSM. lepton MDMs are related to the real parts of the effective couplings, while lepton EDMs are decided by the imaginary ones.

With the lepton MDMs discussed in FIG. \ref{ale}(a),(b), FIG. \ref{alu}(a),(b) and FIG. \ref{aldlu}(a1),(a2),(a3), it is easy to see that the new parameters in the EBLMSSM can solve the problem of opposite symbols of electron MDM and muon MDM. It is worth noting that the absolute ratio of the electron two-loop diagrams' correction to the one-loop contribution and the muon one is approximately $m_e/m_{\mu}$, which is $|\frac{(\Delta a_e^{two-loop}-\Delta a_e^{one-loop})/\Delta a_e^{one-loop}}{{(\Delta a_{\mu}^{two-loop}-\Delta a_{\mu}^{one-loop})/\Delta a_{\mu}^{one-loop}}}|\sim\frac{m_e}{m_{\mu}}$. Besides, the lepton MDMs possesses a large change with $\tan\beta$. Not only that, the lepton MDMs are decoupling with the enlarging $S_m$. In addition, as the introduced parameters in the EBLMSSM, $\mu_{NL}$, $A_E$, $\mu_Y$ and $M_{\tilde{E}}$ can arouse a pretty obvious fluctuation for the lepton MDMs, whose contributions derive the exotic slepton-$\tilde{Y}$ and exotic lepton-$Y$ diagrams.

The lepton EDMs are affected by the CP-violating phases $\theta_{\mu}$, $\theta_{1}$, $\theta_{2}$, $\theta_{M_L}$, $\theta_{\mu_L}$, $\theta_{NL}$, $\theta_{\mu_Y}$ and $\theta_{B_Y}$. Among them, $\theta_{NL}$, $\theta_{\mu_Y}$ and $\theta_{B_Y}$ are the new introduced ones, which influence the numerical results through the exotic slepton-$\tilde{Y}$ and exotic lepton-$Y$ diagrams. As the coupling coefficients of the lepton exotic slepton-$\tilde{Y}$ and lepton exotic lepton-$Y$ vertices, parameters $L_S$ and $L_s$ affect the electron EDM and muon (tau) EDM respectively. We take $L_S$ around 0.001, while $L_s=1$ in our numerical analyses. Electron EDM possesses strict constrains for the EBLMSSM parameter space due to its tiny experimental upper bound, which is $|d_e|<8.7\times10^{-29}$ e.cm. On the basis of considering the cancellation mechanism, we discuss the influence of the new CP-violating phases on electron EDM and find that the numerical results agree well with the experimental upper limit with $0<\theta_{B_Y}<2\pi$, $0<\theta_{\mu_L}<2\pi$ and $\theta_{\mu_Y}$ around $0\sim\pm0.3\pi$. Furthermore, the CP-violating phases $\theta_{\mu}$, $\theta_{1}$, $\theta_{NL}$ and $\tan\beta$ also affect the lepton EDMs obviously. With the development of technology, the lepton EDM may be detected in the near future.

{\bf Acknowledgments}

We are very grateful to Tian-jun Li the teacher of Institute of Theoretical Physics, Chinese Academy of Sciences, for giving us
some useful discussions. This work is supported by the Major Project of National Natural Science Foundation of China (NNSFC) (No. 11535002, No. 11605037, No. 11705045),
the Natural Science Foundation of Hebei province with Grant
No. A2016201010 and No. A2016201069, Hebei Key Lab of Optic-Electronic Information and
Materials, the midwest universities comprehensive strength
promotion project and the youth top-notch talent support program of the Hebei Province.

\appendix
\section{Two-loop corrections to lepton MDMs and EDMs}
Under the assumption $m_F=m_{F_1}=m_{F_2}\gg m_h$, the two-loop Barr-Zee type diagrams contributing to the lepton MDMs and EDMs corresponding to FIG.~\ref{fig2} (b) and (c) can be simplify as
\begin{eqnarray}
&&a_l^{\gamma h}=\frac{G_FQ_fQ_{F_1}m_lm_W^2s_W^2}{16\pi^4}\sum_{F_1=F_2=\chi^\pm,L'}\frac{1}{m_{F_1}}
\Re(H_{hF_1F_2}^L)\Big[1+\ln\frac{m_{F_1}^2}{m_h^2}\Big],
\nonumber\\&&d_l^{\gamma h}=\frac{eG_FQ_fQ_{F_1}m_W^2s_W^2}{32\pi^4}\sum_{F_1=F_2=\chi^\pm,L'}\frac{1}{m_{F_1}}
\Im(H_{hF_1F_2}^L)\Big[1+\ln\frac{m_{F_1}^2}{m_h^2}\Big],
\nonumber\\&&a_l^{Zh}=-\frac{G_F m_l m_W^2s_W}{64e\pi^4c_W}\sum_{F_1=F_2=\chi^{\pm},L',\chi^0,N'}\frac{H_{hl\bar{l}}}{m_{F_1}}\Big[\varrho_{1,1}(m_Z^2,m_h^2)-\ln{m_{F_1}^2}-1\Big]
\nonumber\\&&\qquad\quad\times(T_f^Z-2Q_{F_1}s_W^2)\Re(H_{hF_1F_2}^LH_{ZF_1F_2}^L+H_{hF_1F_2}^RH_{ZF_1F_2}^R)
\nonumber\\&&d_l^{Zh}=-\frac{G_F m_W^2s_W}{64\pi^4c_W}\sum_{F_1=F_2=\chi^{\pm},L',\chi^0,N'}\frac{H_{hl\bar{l}}}{m_{F_1}}\Big[\varrho_{1,1}(m_Z^2,m_h^2)-\ln{m_{F_1}^2}-1\Big]
\nonumber\\&&\qquad\quad\times(T_f^Z-2Q_{F_1}s_W^2)\Im(H_{hF_1F_2}^LH_{ZF_1F_2}^L-H_{hF_1F_2}^RH_{ZF_1F_2}^R).
\end{eqnarray}

Under the assumptions $m_F=m_{F_1}=m_{F_2}\gg m_W$ and $m_F=m_{F_1}=m_{F_2}\gg m_Z$, the two-loop Rainbow type diagrams contributing to the lepton MDMs and EDMs corresponding to FIG.~\ref{fig2} (d), (e) and (f) can be simplify as
\begin{eqnarray}
&&a_l^{WW}=\frac{G_F m_l^2}{192\sqrt{2}\pi^4}\sum_{F_1=\chi^{\pm},L'}\sum_{F_2=\chi^0,N'}\Big\{(18Q_{F_1}-13)(|H_{WF_1F_2}^L|^2+|H_{WF_1F_2}^R|^2)
\nonumber\\
&&\qquad\quad+3(Q_{F_1}-3)(|H_{WF_1F_2}^L|^2-|H_{WF_1F_2}^R|^2)+11\Re(H_{WF_1F_2}^{R*}H_{WF_1F_2}^L)\Big\},
\nonumber\\&&d_l^{WW}=-\frac{G_F em_l}{64\sqrt{2}\pi^4}\sum_{F_1=\chi^{\pm},L'}\sum_{F_2=\chi^0,N'}(2+Q_{F_1})\Im(H_{WF_1F_2}^{R*}H_{WF_1F_2}^L),
\nonumber\\&&a_l^{\gamma\gamma}=\frac{\sqrt{2}e^2G_FQ_{F_1}^2m_l^2}{180\pi^4}\sum_{F_1=F_2=\chi^\pm,L'}\frac{m_W^2}{m_{F_1}^2},
\nonumber\\&&a_l^{\gamma Z}=\frac{eG_FQ_fQ_{F_1}m_l^2m_W^2s_W}{64\sqrt{2}\pi^4c_W}\sum_{F_1=F_2=\chi^\pm,L'}\frac{1}{m_{F_1}^2}
\Re(H_{ZF_1F_2}^L-H_{ZF_1F_2}^R)\Big[35+\ln\frac{m_{F_1}^2}{m_Z^2}\Big],
\nonumber\\&&d_l^{\gamma\gamma}=d_l^{\gamma Z}=0.
\end{eqnarray}
$H_{hF_1F_2}^{L,R}$, $H_{ZF_1F_2}^{L,R}$ and $H_{hl\bar{l}}$ represent the corresponding coupling coefficients.

 \end{document}